\documentclass[prd,preprint,nofootinbib]{revtex4-1} 

\usepackage{latexsym,verbatim,enumerate}
\usepackage{amssymb}
\usepackage{amsfonts}
\usepackage{amsmath,color}
\usepackage{graphicx}
\usepackage{bm}
\usepackage[utf8]{inputenc}
\usepackage[T1]{fontenc}

\usepackage{hyperref}

\hypersetup{
    colorlinks=true,
    linkcolor=red,
    citecolor=green,
    filecolor=magenta,      
    urlcolor=cyan,
    pdftitle={Gravitational lensing observables in stationary and axisymmetric solutions in general relativity},
    pdfpagemode=FullScreen,}

\def\mb#1{\mathbf{#1}}

\def\ber{\begin{eqnarray}}
\def\eer{\end{eqnarray}}
\def\beq{\begin{equation}}
\def\eeq{\end{equation}}

\def\ed{\end{document}}

\def\dla#1{\frac{\mathrm{d} #1}{\mathrm{d}\lambda}}

\makeatletter
\let\jnl@style=\rm
\def\ref@jnl#1{{\jnl@style#1}}

\def\aj{\ref@jnl{AJ}}                   
\def\actaa{\ref@jnl{Acta Astron.}}      
\def\araa{\ref@jnl{ARA\&A}}             
\def\apj{\ref@jnl{ApJ}}                 
\def\apjl{\ref@jnl{ApJ}}                
\def\apjs{\ref@jnl{ApJS}}               
\def\ao{\ref@jnl{Appl.~Opt.}}           
\def\apss{\ref@jnl{Ap\&SS}}             
\def\aap{\ref@jnl{A\&A}}                
\def\aapr{\ref@jnl{A\&A~Rev.}}          
\def\aaps{\ref@jnl{A\&AS}}              
\def\azh{\ref@jnl{AZh}}                 
\def\baas{\ref@jnl{BAAS}}               
\def\bac{\ref@jnl{Bull. astr. Inst. Czechosl.}}
\def\caa{\ref@jnl{Chinese Astron. Astrophys.}}
\def\cjaa{\ref@jnl{Chinese J. Astron. Astrophys.}}
\def\icarus{\ref@jnl{Icarus}}           
\def\jcap{\ref@jnl{J. Cosmology Astropart. Phys.}}
\def\jrasc{\ref@jnl{JRASC}}             
\def\memras{\ref@jnl{MmRAS}}            
\def\mnras{\ref@jnl{MNRAS}}             
\def\na{\ref@jnl{New A}}                
\def\nar{\ref@jnl{New A Rev.}}          
\def\pra{\ref@jnl{Phys.~Rev.~A}}        
\def\prb{\ref@jnl{Phys.~Rev.~B}}        
\def\prc{\ref@jnl{Phys.~Rev.~C}}        
\def\prd{\ref@jnl{Phys.~Rev.~D}}        
\def\pre{\ref@jnl{Phys.~Rev.~E}}        
\def\prl{\ref@jnl{Phys.~Rev.~Lett.}}    
\def\pasa{\ref@jnl{PASA}}               
\def\pasp{\ref@jnl{PASP}}               
\def\pasj{\ref@jnl{PASJ}}               
\def\rmxaa{\ref@jnl{Rev. Mexicana Astron. Astrofis.}}%
\def\qjras{\ref@jnl{QJRAS}}             
\def\skytel{\ref@jnl{S\&T}}             
\def\solphys{\ref@jnl{Sol.~Phys.}}      
\def\sovast{\ref@jnl{Soviet~Ast.}}      
\def\ssr{\ref@jnl{Space~Sci.~Rev.}}     
\def\zap{\ref@jnl{ZAp}}                 
\def\nat{\ref@jnl{Nature}}              
\def\iaucirc{\ref@jnl{IAU~Circ.}}       
\def\aplett{\ref@jnl{Astrophys.~Lett.}} 
\def\apspr{\ref@jnl{Astrophys.~Space~Phys.~Res.}}
\def\bain{\ref@jnl{Bull.~Astron.~Inst.~Netherlands}}
\def\fcp{\ref@jnl{Fund.~Cosmic~Phys.}}  
\def\gca{\ref@jnl{Geochim.~Cosmochim.~Acta}}   
\def\grl{\ref@jnl{Geophys.~Res.~Lett.}} 
\def\jcp{\ref@jnl{J.~Chem.~Phys.}}      
\def\jgr{\ref@jnl{J.~Geophys.~Res.}}    
\def\jqsrt{\ref@jnl{J.~Quant.~Spec.~Radiat.~Transf.}}
\def\memsai{\ref@jnl{Mem.~Soc.~Astron.~Italiana}}
\def\nphysa{\ref@jnl{Nucl.~Phys.~A}}   
\def\physrep{\ref@jnl{Phys.~Rep.}}   
\def\physscr{\ref@jnl{Phys.~Scr}}   
\def\planss{\ref@jnl{Planet.~Space~Sci.}}   
\def\procspie{\ref@jnl{Proc.~SPIE}}   

\makeatother

\begin{document}

\author{Matteo Luca Ruggiero}
\email{matteoluca.ruggiero@unito.it}
\affiliation{Dipartimento di Matematica ``G.Peano'', Universit\`a degli studi di Torino, Via Carlo Alberto 10, 10123 Torino, Italy}
\affiliation{INFN - LNL , Viale dell'Universit\`a 2, 35020 Legnaro (PD), Italy}

\author{Davide Astesiano}
\email{davide.astesiano@venturilab.ch}
\affiliation{Science Institute, University of Iceland,
Dunhaga 3, 107 , Reykjav\'{\i}k, Iceland}
\affiliation{Mathematics Division, Venturi Space, Route du Pâqui 1, 1720 Corminboeuf, Switzerland}

\date{\today}

\title{Gravitational lensing observables in stationary and axisymmetric solutions in general relativity}

\begin{abstract}
We investigate light propagation in self-gravitating systems composed of an axially symmetric, stationary, rotating dust fluid. These configurations are intrinsically relativistic, sustained entirely by their rotation, since no compact or finite dust distribution can exist under the same symmetry conditions in Newtonian gravity. In such systems, rotational effects arise from off-diagonal components of the spacetime metric, which are not negligible compared to their Newtonian counterparts. We analyze how these components affect the deflection angle of light, showing that they can be interpreted as contributing an additional effective mass. Moreover, their presence can, in principle, be detected through the characteristic asymmetry they induce in the images of background sources.
\end{abstract}

\maketitle

\section{Introduction }\label{sec:intro}

Einstein's theory of gravitation, General Relativity (GR), profoundly changed our vision of the Universe by  shifting the paradigm from the  Newtonian concept of gravity as action at a distance to  a geometric interpretation in which gravitation arises from the curvature of four-dimensional spacetime.  Since its publication, GR has passed many tests with flying colors, ranging from experiments conducted within the Solar System to observations in deep space \cite{willrev}. GR is not only an extension of classical gravitation, because it predicted entirely new phenomena without  a Newtonian analogue, which have been successfully detected, such as neutron stars, black holes and gravitational waves. These phenomena, however, typically arise in regions where spacetime is highly deformed. On the other hand, when the gravitational field is weak and can be treated as a perturbation of Minkowski flat spacetime, GR effects are expected to be small corrections with respect to Newtonian gravity. However, even when the gravitational field is weak, there are general relativistic effects that have no counterpart in classical gravitation. For example,  when Einstein's equations are solved in the weak-field and slow-motion approximation, the general solution contains gravitomagnetic terms originating from the rotation of the sources, i.e. on their mass currents. As a result, a well know gravitoelectromagnetic analogy can be established and used to describe new effect in terms of known electromagnetic ones  \cite{Ruggiero:2002hz,mashhoon03}. In this framework, gravitomagnetic  effects are extremely small and their detection requires highly accurate measurement strategies \cite{Ruggiero:2023ker}. 

The situation can radically change if we consider self-gravitating systems composed of an axially symmetric and stationary rotating dust fluid. For such configurations,  there are exact solutions of Einstein's equations which were thoroughly studied from a physical and mathematical perspective \cite{Geroch:1970nt,Geroch:1972yt,HansenWinicour1,HansenWinicour2,Winicour,stephani_kramer_maccallum_hoenselaers_herlt_2003}. These systems are unique in that they exist exclusively in the framework of GR: indeed, no compact or finite dust configuration can exist in Newtonian gravity in the given symmetry conditions \cite{bonnor1977rotating,Ruggiero:2021lpf}; notably, their existence relies on rotation effects which are of the same order of magnitude of Newtonian ones \cite{Ruggiero:2023geh}.  The low-energy limits of these solutions were analyzed in our previous works \cite{Astesiano:2022ozl, Ruggiero:2023geh, Astesiano:2024zzz}, with particular emphasis on the dynamics of massive particles. It was shown that the velocity profiles of the sources can become asymptotically flat, in contrast to Newtonian predictions. The aim of this paper is to focus on the dynamics of massless particles, in order to investigate whether distinctive gravitational lensing observables can be associated with these systems.

\section{Gravitational Lensing in GR }\label{sec:lensingGR}

The propagation of light rays in curved spacetime can be derived from a suitable formulation of the Fermat's principle, which states that a light ray traveling from a source to an observer follows a trajectory for which the arrival time is stationary, meaning that, to first order, small variations in the path do not change the travel time  \cite{1990ApJ...351..114K,ehlers,perlick2000ray}. In particular, we are interested in the application of the Fermat's principle to the study of propagation of light rays in stationary spacetimes. To this end, we consider an asymptotically flat spacetime, described by the metric\footnote{Latin indices run from 1 to 3, and refer to space components, while Greek indices run from 0 to 3, and label spacetime components; boldface letter such as $\mb w$ refers to three-vectors; the spacetime signature is $(-1,1,1,1)$}
\beq
ds^{2}=g_{00}(x^{i})c^{2}dt^{2}+2g_{0i}(x^{i})cdt dx^{i}+g_{ij}(x^{i})dx^{i}dx^{j}, \label{eq:metricastat}
\eeq
in coordinates $\left(ct,x^{i} \right)$. The  spacetime interval (\ref{eq:metricastat}) can be written in the form \cite{Ruggiero:2023ker}
\beq
ds^{2}=-c^{2}dT^{2}+d\sigma^{2}, \label{eq:ds2splitting}
\eeq
setting $\displaystyle dT=-u_{\alpha}dx^{\alpha}, \quad d\sigma^{2}=\gamma_{ij}dx^{i}dx^{j}$, where $\displaystyle u_{\alpha}=\frac{g_{0\alpha}}{\sqrt{-g_{00}}}, \quad \gamma_{\alpha\beta}=g_{\alpha\beta}+u_{\alpha}u_{\beta}$.  Hence, since $ds^{2}=0$ for a light ray, on a future-directed light curve, we obtain
\beq
cdt=\frac{1}{\sqrt{-g_{00}}}d\sigma-\frac{g_{0i}}{g_{00}}dx^{i}. \label{eq:fermatdt1}
\eeq
We study a light ray propagating in the spacetime (\ref{eq:metricastat}); without loss of generality, we may suppose that $t=0$ at the emission. We consider  stationary observers: their  world-line is $x^{i}=\mathrm{const}$, parametrized with the coordinate time $t$. A light ray propagates along a null curve; we are interested in its spatial projection, denoted as $\tilde{\gamma}$, which we parameterize by $\lambda$ and describe with the functions $x^{i}(\lambda)$. The arrival time can then be expressed as:
\beq
t=\frac 1 c \int_{\tilde{\gamma}}\left( \frac{1}{\sqrt{-g_{00}}}d\sigma-\frac{g_{0i}}{g_{00}}dx^{i} \right). \label{eq:proptime1}
\eeq
Accordingly, the Fermat's principle is equivalent to saying that $\delta t=0$, which can be written in the form 
\beq
\delta \int_{\tilde \gamma} n d\sigma=0, \label{eq:fermat2}
\eeq
where $\displaystyle n=\left( \frac{1}{\sqrt{-g_{00}}}-\frac{g_{0i}}{g_{00}}\frac{dx^{i}}{d\sigma} \right) $ is an effective refraction index, which depends both on position and the propagation direction, and $d\sigma$ is the spatial arc length \cite{serenoPRD2003}. Fermat's principle allows us to detetermine the spatial path of light rays and, consequently, the deflection angle. To this end, setting $\displaystyle \dot x^{i}=\dla{x^{i}}$, the Fermat's principle can be regarded as a variational principle $\displaystyle \delta \int n(x^{i},\dot x^{i}) d\lambda=0$ for the Lagrangian
\beq
n(x^{i},\dot x^{i})=\left( \frac{1}{\sqrt{-g_{00}(x^{i})}}\sqrt{\gamma_{ij}(x^{i})\dot x^{i}\dot x^{j}}-\frac{g_{0i}(x^{i})}{g_{00}(x^{i})}\dot x^{i}\right). \label{eq:ferEL2}
\eeq
from which, thanks to the Euler-Lagrange equations, it is possible to obtain the functions  $x^{i}(\lambda)$ which determine the spatial path of light rays.
If $G$ denotes the gravitational constant, we know that calculating the contribution of order $G^{N}$ to the lensing quantities requires knowledge of the deflected light ray path up  to the order $G^{N-1}$ \cite{fischbach}; consequently, at the lowest meaningful order applicable to realistic astrophysical conditions, light deflection represents only a small perturbation to the unperturbed Euclidean path. Hence,  $d\lambda$ can be taken as the Euclidean arc length, i.e. in Cartesian coordinates $d\lambda=\sqrt{\delta_{ij}dx^{i}dx^{j}}$, so that $\displaystyle \sqrt{\delta_{ij}\dot x^{i}\dot x^{j}}=1$.
Using suitable isotropic coordinates, $\gamma_{ij}(x^{i})=f(x^{i})\delta_{ij}$, so that Eq. (\ref{eq:ferEL2}) becomes $\displaystyle n(x^{i},\dot x^{i})=n_{0}(x^{i})\sqrt{\delta_{ij}\dot x^{i}\dot x^{j}}+w_{i}(x^{i})\dot x^{i}$, where
\beq
n_{0}(x^{i}) = \frac{f(x^{i})}{\sqrt{-g_{00}(x^{i})}},  \quad w_{i}(x^{i})= -\frac{g_{0i}(x^{i})}{g_{00}(x^{i})}.  \label{eq:ferEL5}
\eeq
Using a three-dimensional vector notation, $\displaystyle \mathbf{v}=\dla{\mathbf x}$ denotes the unit vector tangent to the spatial path of the light ray, and $\mathbf w$ is the vector whose components are $w_{i}$. Then, the Euler-Lagrange equations read \cite{ehlers} 
\beq
n_{0} \dla{\mathbf v}=\bm \nabla_{\bot} n_{0}+\mb v \wedge \bm \nabla \wedge \mb w, \label{eq:ferEL6}
\eeq
where $\displaystyle \bm \nabla_{\bot} n_{0}=\bm \nabla n_{0}-\mb v \left( \mb v \cdot \bm \nabla n_{0} \right)$ is the projection of $\bm \nabla n_{0}$ onto the plane orthogonal to propagation direction of the light ray.  The deflection angle is defined as the difference between the initial and final ray direction, $\displaystyle \bm \alpha=\mb v_{in}-\mb v_{fin}$, so that  from Eq. (\ref{eq:ferEL6}) we have
\beq
\bm \alpha = \int_{u.p.} n_{0} ^{-1} \left( \bm \nabla_{\bot} n_{0}+\mb v \wedge \bm \nabla \wedge \mb w \right) d\lambda, \label{eq:defalpha2}
\eeq
where the integral is taken over the straight unperturbed path (u.p.).

\section{The Weak Field Solution} \label{sec:hyp}

We consider  a physical model consisting of a stationary, axially symmetric rotating dust fluid. Owing to the symmetries of the system, the spacetime admits one timelike and one spacelike Killing vector, which allows for an exact solution of Einstein's equations. From a physical standpoint, the characteristics of a solution are determined by the velocity and density profiles of the sources. However, it is crucial to highlight that for this class of solutions, even when the velocity and density profiles are specified, the solution is not directly determined: there remain certain degrees of freedom, as discussed in the work by \citet{Astesiano:2021ren}. In particular, we are concerned with the  the low-energy limits of these solutions: this amounts to considering them in the weak-field regime for the gravitational field  \cite{Astesiano:2022ozl,Ruggiero:2023geh,Astesiano:2024zzz} where, using cylindrical coordinates $(x^0=ct,r,z,\phi)$, the solution at the leading order is
\begin{align}
     ds^2&= -c^2\left(1-\frac{2\Phi}{c^2}-\frac{\psi^2}{r^2c^2}\right)dt^2-2 \frac{\psi}{c} c dt d\phi +r^2\left(1+ \frac{2\Phi}{c^2}\right)d\phi^2+ e^{\Psi} \left(dr^2+dz^2\right), \label{eq:metricsol} 
\end{align}
and   the function $e^{\Psi}$ is determined by the line integrals
\begin{align}
    \Psi_{,r}=\frac{1}{2r} \left[2r \partial_r\left(\frac{2\Phi}{c^2}+ \frac{\psi^2}{c^2r^2}\right)+ \frac{\psi^2_{,z}-\psi^2_{,r}}{c^2} \right], \label{eq:defPSIr}
\end{align}
or, equivalently,
\begin{align}
 \Psi_{,z}=\frac{1}{2r} \left[2r \partial_z\left(\frac{2\Phi}{c^2}+ \frac{\psi^2}{c^2r^2}\right)-\frac{2}{c^2} \psi_{,r} \psi_{,z}\right]. \label{eq:defPSIz}
\end{align}
We remark that, due to the symmetry of the system, all functions depend only on $(r,z)$. In the above equations, $\Phi$ is the generalization of the Newtonian potential, and it is the solution of the {modified} Poisson equation
\beq
\nabla^2 \Phi+ \frac{(\partial_{z}\psi)^2+(\partial_r\psi-2 \frac{\psi}{r})^2}{2 r^2}=-4\pi G\rho. \label{eq:poisson}
\eeq
The function $\psi$  determines the dragging of the inertial frames \cite{Astesiano:2022ozl}, so we will refer to it as \textit{dragging term} and it is the solution of the homogenous Grad-Shafranov equation \cite{grad1958hydromagnetic,shafranov1958magnetohydrodynamical}
\beq
\partial_{rr} \psi+ \partial_{zz} \psi- \frac{\partial_r \psi}{r}=0. \label{eq:GS1}
\eeq
{We remark that, as discussed by \cite{Astesiano:2022gph} and \citet{Galoppo:2024ttc}, $\psi$ represents the \textit{quasilocal angular momentum} per unit of mass associated with
spacetime rotation.}
As we said before, we see that once the sources are given, the gravitational field is not uniquely determined, since it depends on the particular solution $\psi$ of the Grad-Shafranov equation (\ref{eq:GS1}). Remarkably, this function modifies the interplay between the sources of the gravitational field and the Newtonian potential $\Phi$, with an effective matter density in the form
\beq
\rho_{\psi}=\frac{1}{4\pi G}\left(\frac{(\partial_{z}\psi)^2+(\partial_r\psi-2 \frac{\psi}{r})^2}{2 r^2} \right).  \label{eq:rhopsi1}
\eeq
{Equation (\ref{eq:poisson}) generalizes the Poisson equation to the general relativistic regime, reducing to it when $\psi$ vanishes. The extra term in equation (\ref{eq:rhopsi1}) reflects rotational energy from the quasilocal angular momentum of the averaged background, a feature absent in the Newtonian approximation with a Minkowski background  \cite{Galoppo:2024ttc}; we notice that it is independent of the actual matter content of the system, as described by $\rho$.}

 It is crucial to emphasize  that the solutions for the dragging term $\psi$ are not necessarily negligible compared to classical Newtonian effects \cite{Astesiano:2024zzz}. This fact, can be easily understood if we look at the $z$ component of the geodesic equations describing the motion of the sources of the gravitational field
 \beq
  \frac{V}{r} \partial_z \psi=\partial_z \Phi+ \frac{\psi}{r^2} \partial_z \psi, \label{SGMzz}\\
 \eeq
where $V$ is the dust velocity as measured by asymptotic inertial observers. Equation (\ref{SGMzz}) implies that, to maintain equilibrium along the axis of symmetry, the rotational effects, associated with the off-diagonal components of the spacetime metric and represented by $\psi$, must be significant when compared to the Newtonian effects denoted by $\Phi$. This implies that the system is dominated by rotational effects that are comparable in magnitude to the Newtonian ones. Therefore, it constitutes a genuinely relativistic configuration that cannot be explained using Newtonian analogies, but only within the framework of General Relativity \cite{Ruggiero:2023geh}.

\section{Light deflection in the axially symmetric and stationary solution} \label{sec:ldasss}

%

We apply the formalism described in Section \ref{sec:lensingGR} to the weak-field solution (\ref{eq:metricsol}). We start with some preliminary considerations, deriving by a closer inspection of the deflection angle formula (\ref{eq:defalpha2}) which, to begin with, to lowest approximation order can be written as
\beq
\bm \alpha = \int_{u.p.}  \left( \bm \nabla_{\bot} n_{0}+\mb v \wedge \bm \nabla \wedge \mb w \right) d\lambda \label{eq:defalpha20}
\eeq
i.e. neglecting the $n_{0} ^{-1}$ term in the integral. Under standard conditions, such as those associated with the classical weak-field solution that gives rise to the gravitoelectromagnetic analogy \cite{Ruggiero:2023ker, mashhoon03, Ruggiero:2002hz}, the first term in the integral (\ref{eq:defalpha20}) has a gravitoelectric origin: specifically, this term represents a contribution arising from the Newtonian potential, which appears in the diagonal components of the spacetime metric. However, when we consider the spacetime element (\ref{eq:metricsol}), we se that this is not the case. In fact, the dragging term $\psi$ (i) appears in the diagonal elements of the spacetime metric, and (ii) enters the definition of the spatial metric $\gamma_{ij}$. In both cases, the resulting effects are quadratic, that is, proportional to $\psi^{2}$, and their contribution to the deflection angle is of the order $O(c^{-2})$.   On the other hand, the second term in the integral (\ref{eq:defalpha20}) is purely gravitomagnetic, as it arises from the off-diagonal components of the metric, which are linear in $\psi$. Consequently, we find that, unlike in the classical weak-field solution, where the gravitomagnetic contribution to gravitational light deflection is of the order $O(c^{-3})$ \cite{ehlers, serenoPRD2003}, in this case, it is of the order $O(c^{-1})$. This aspect of these solutions to Einstein's equations has not been emphasized previously, even though \citet{galoppo2,galoppo1} developed  an approach to gravitational lensing based on  the Gauss-Bonnet theorem, which can be applied  only for light rays propagating in the $z=0$ plane. Here we aim to focus on these peculiarities, by explicitly calculating the gravitomagnetic contribution
\beq
\bm \alpha_{GM} =\int_{u.p.} \mb v \wedge \bm \nabla \wedge \mb w\,  d\lambda. \label{eq:defalphaGM1}
\eeq
 to the deflection angle, which is the leading term determined by $\psi$, as we have seen that the contribution of the dragging term to the first term in (\ref{eq:defalpha20}) is quadratic.
 We remark that, in this class of solutions of Einstein's equations, To explicitly calculate (\ref{eq:defalphaGM1}) we must chose a solution of the homegenous Grad-Shafranov equation (\ref{eq:GS1}), which we take  in the form
 \beq
\psi=\frac{\kappa r^{2}}{\left(z^{2}+r^{2}\right)^{3/2}} \label{eq:psidef1}
\eeq
{The choice of this solution is motivated by the fact that it corresponds to the dipole-like term present in  the weak-field expansion of the Kerr’s solution \cite{Ruggiero:2023geh,Astesiano:2024zzz}.} We notice that since $\psi$ has the dimension of  $\mathrm{length} \times \mathrm{velocity}$, $\kappa$ has the dimension of a $\mathrm{length^{2}\times velocity}$. Using the  expression of $\psi$ given in Eq. (\ref{eq:psidef1}), and taking into account that $\displaystyle \mb w=w_{\phi}\mb u_{\phi}=\frac{\psi}{cr} \mb u_{\phi}$ at the leading order, we may define the gravitomagnetic field $\displaystyle \mb B= \bm \nabla \wedge \mb w$ whose components are conveniently expressed in Cartesian coordinates
\beq
\mb B = \frac{\kappa}{c}\frac{3zx}{\left(x^{2}+y^{2}+z^{2} \right)^{5/2}}\mb u_{x}+\frac{\kappa}{c}\frac{3zy}{\left(x^{2}+y^{2}+z^{2} \right)^{5/2}}\mb u_{y}-\frac{\kappa}{c}\frac{\left(r^{2}-2z^{2} \right)}{\left(r^{2}+z^{2} \right)^{5/2}} \mb u_{z} \label{eq:defBGMcart}
\eeq 

\begin{figure}[h]
\begin{center}
\includegraphics[scale=.40]{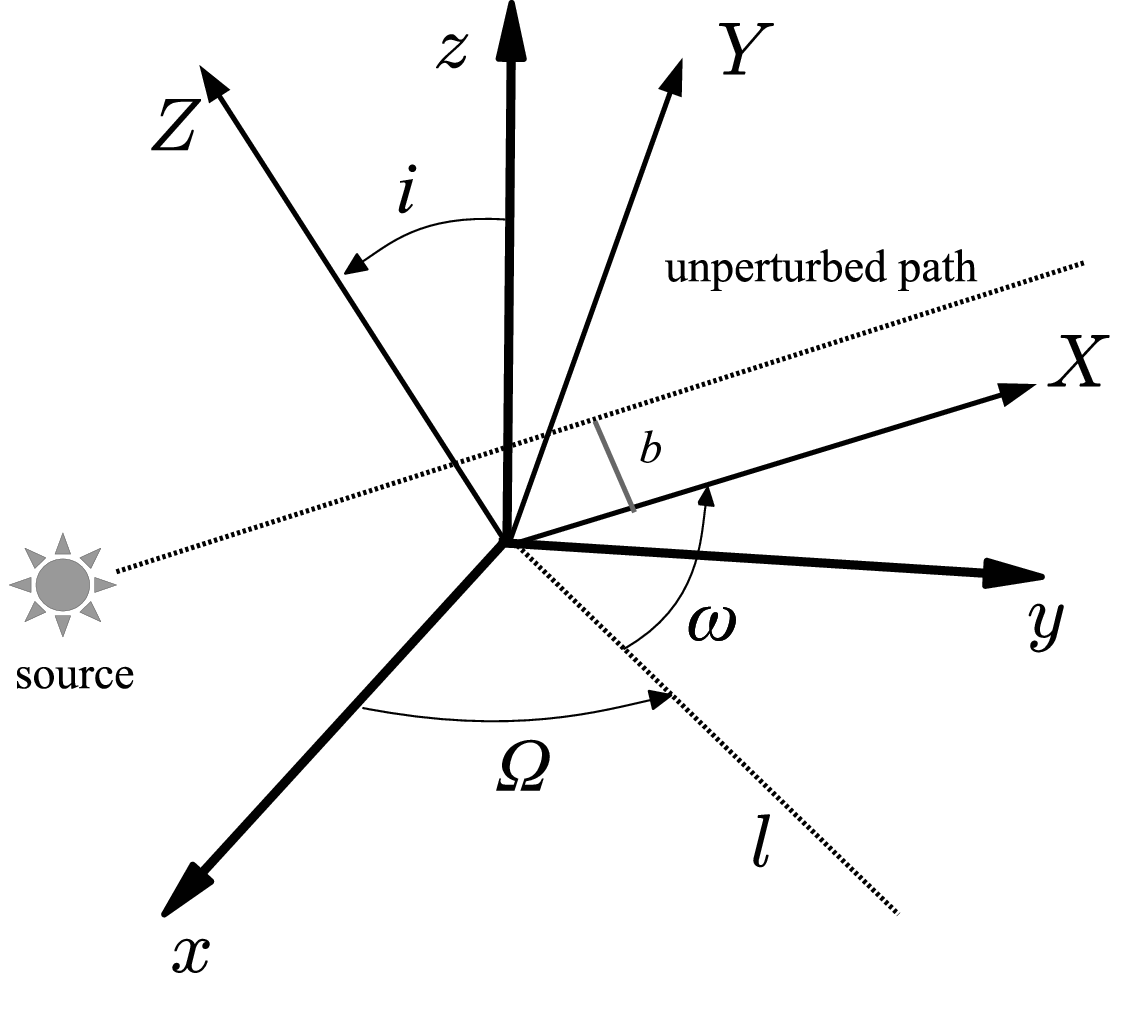}
\caption{The unperturbed path   lies in the $XY$ plane. We denote with $\Omega$  the angle between the $x$ axis and $l$, which is the intersection between the propagation plane and the reference plane $xy$, while the angle between the $z$ and $Z$ axes is $i$; eventually, we denote by $\omega$  the angle between $l$  and the $X$ axis.} \label{fig:fig1}
\end{center}
\end{figure}  

To define the geometry of lensing in arbitrary conditions, we consider a Cartesian reference system $XYZ$, centered at $r=0,z=0$, such that unperturbed path of light lies in the $XY$ plane, which is then defined  by the closest approach position and the propagation direction. In particular, we suppose the unperturbed path is $\mb x(\lambda)=\lambda \mb u_{X}+b\mb u_{Y}$, where $b$ defines the closest approach position,
while $\mb v= \mb u_{X}$. The relation between this Cartesian reference system and the Cartesian system $xyz$  adapted to the symmetries of the source is described in Fig. \ref{fig:fig1}, and the following relations hold between the unit vectors of the two Cartesian coordinate systems \cite{bertotti2012physics}:
{
\begin{eqnarray}
\mb u_{X} & = & \left(\cos \omega \cos \Omega-\sin \omega \cos i \sin \Omega \right) \mb u_{x}+\left(\cos \omega \sin \Omega+\sin \omega \cos i \cos \Omega \right) \mb u_{y}+\sin \omega \sin i  \, \mb u_{z} \nonumber \\
\mb u_{Y} & = & \left(-\sin \omega \cos \Omega-\cos \omega \cos i \sin \Omega \right) \mb u_{x}+\left(-\sin \omega \sin \Omega+\cos \omega \cos i \cos \Omega \right) \mb u_{y}+\cos \omega \sin i  \, \mb u_{z} \nonumber \\
\mb u_{Z}&=& \sin i \sin \Omega \, \mb u_{x}-\sin i \cos \Omega \, \mb u_{y}+ \cos i \mb u_{z} \label{eq:trasxyzXYZ} 
\end{eqnarray}}

By integrating  $\displaystyle \bm \alpha_{GM} =\int_{u.p.} \mb v \wedge \mb B\, d\lambda$ along the unperturbed path, we obtain the components of the deflection angle
\begin{eqnarray}
\alpha_{GM,x}&=&-\frac{2\kappa}{cb^{2}}\left[\sin \Omega \cos \omega \left(2\cos^{2}i-1 \right)+\sin \omega \cos i \cos \Omega \right] \label{eq:alphaGMx1}\\
\alpha_{GM,y}&=&-\frac{2\kappa}{cb^{2}}\left[\cos \Omega \cos \omega \left(-2\cos^{2}i+1 \right)+\sin \omega \cos i \sin \Omega \right] \label{eq:alphaGMy1}\\
\alpha_{GM,z}&=&\frac{4\kappa}{cb^{2}}\left[\cos \omega \sin i \cos i \right] \label{eq:alphaGMz1}
\end{eqnarray}	
The case of the $z$ axis perpendicular to the line of sight, is obtained by setting $i=\Omega=\omega=0$, and we get
\beq
\displaystyle \bm \alpha = \frac{2\kappa}{cb^{2}}\mb u_{y} \label{eq:alphatot3}
\eeq

We can calculate the effect for a general solution of the Grad-Shafranov equation (\ref{eq:GS1}); for simplicity, we consider the case of the $z$ axis perpendicular to the line of sight. In particular, it is possible to show \cite{Astesiano:2024zzz} that the solution to Eq. (\ref{eq:GS1}) is given by
\beq
\psi = \int_{0}^{\infty} d\xi \, C(\xi) \, \frac{r^2}{[(z+\xi)^2 + r^2]^{3/2}}.
\eeq
Letting $r^2= x^2+b^2$ and integrating over $x$ from $-\infty$ a $\infty$, we get
\beq
\bm \alpha_{GM} = \frac 1 c \int_{-\infty}^{+\infty} \left[ 4b\xi \frac{ C(\xi)}{(b^2 + \xi^2)^2} \mb{u}_z 
- 2\frac{(b^2 - \xi^2)}{(b^2 + \xi^2)^2} C(\xi) \, \mb{u}_y \right] d\xi.
\eeq
If we expand the results in power of large $b$, we obtain the result at the first order
\beq
\lim_{b\rightarrow \infty}  \bm \alpha_{GM} = \frac 1 c\int \left[ \frac{4 \xi C(\xi)}{b^3} d\xi + \mathcal{O}\left(\frac{1}{b^5}\right) \right] \mb{u}_z 
+ \int \left[ \frac{2 C(\xi)}{b^2} d\xi + \mathcal{O}\left(\frac{1}{b^4}\right) \right] \mb{u}_y
\eeq
By choosing $C(\xi)= \kappa \delta(\xi)$, we obtain again the result in Eq. (\ref{eq:alphatot3}) at first order.
If $C(\xi)$ is an even function than the integral along the $z$ direction is zero. This shows that when the solution is symmetric for exchange $z \rightarrow-z$, there is not net contribution in this direction.

\begin{figure}[h]
\begin{center}
\includegraphics[scale=.70]{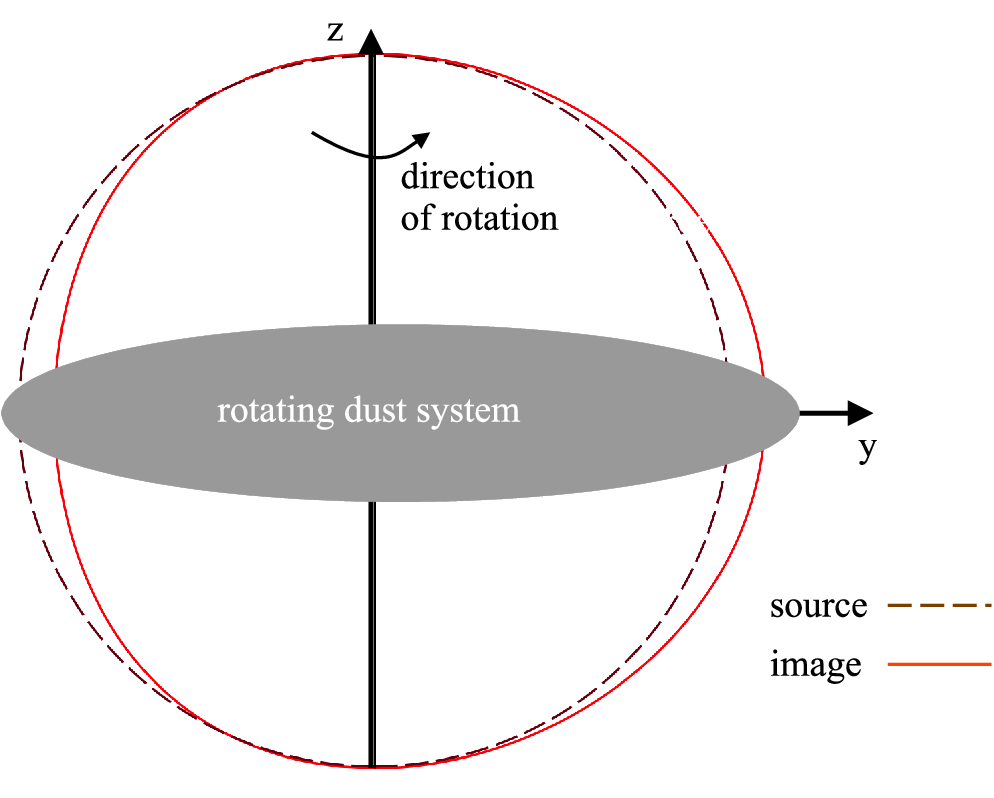}
\caption{{The sources of light rays are distributed along a circle in the plane of the lens; the dust system rotates about the 
$z$ axis; the resulting image exhibits a deformation determined by the gravitomagnetic term.}}  \label{fig:fig2}
\end{center}
\end{figure}

The above expressions can be used to evaluate the impact of the dragging terms once that the orientation of the {rotating system} is known. However, to better understand these effects, we try to give a pictorial description. To this end,  we consider the  lens equation which relates the (angular) position of the images $\bm \theta$ with the  (angular) position of the sources $\bm \beta$, as a function of the deviation angle $\bm \alpha$
\beq
\bm \beta=\bm \theta-\frac{D_{ds}}{D_{s}}\bm \alpha, \label{eq:lens1}
\eeq
where $D_{s}$, $D_{ds}$ are respectively the distances between the observer and the source and between the lens and the source \cite{ehlers}; {in particular, the thin-lens approximation is employed, wherein the distances between the source, lens, and observer are much greater than the characteristic scale of the lensing system, a condition that is well satisfied for astronomical objects observed from Earth. }  Here we take into account only the gravitomagnetic contribution to the deflection angle and,  for the sake of simplicity, we consider the case of the $z$ axis perpendicular to the line of sight, so that the unperturbed path is  $\mb x(\lambda)=\lambda \mb u_{x}+b\mb u_{y}$. Qualitatively, the effect is described in Fig. \ref{fig:fig2}, where we see that {if the source of light rays are distributed along a circle centered in $z=r=0$ in the plane of the lens, the image is  shifted by the presence of the dragging term  $\psi$;}  in particular,  we see that photons traveling in the equatorial plane in the same direction as the sources are deflected toward it, while those moving in the opposite direction are deflected away.

\section{Discussion and conclusions} \label{sec:disconc}

Self-gravitating systems composed of an axially symmetric rotating dust fluid are purely relativistic objects, sustained by their rotation. In these systems, the rotational effects determined by the off-diagonal dragging term $\psi$ in the metric (\ref{eq:metricsol}) cannot be neglected in comparison to the Newtonian contributions described by the potential $\Phi$. In this context, our previous works  \cite{Astesiano:2024zzz} demonstrated that the velocity profile of the sources can become asymptotically flat, which is at odds with respect to Newtonian expectations. Given that flat rotation curves are a key feature in galactic dynamics and a major indication of dark matter, it is unsurprising that such systems have been explored as relativistic models for galaxies \cite{Balasin:2006cg,Cooperstock:1993en,crosta2020testing,Astesiano:2021ren,Astesiano:2022gph,Astesiano:2022ozl,Astesiano:2022ghr,Ruggiero:2023geh,10.1093/mnras/stae855,Re:2024qco,Galoppo:2024ttc,Galoppo:2024mfw}.

{In this context, we would like to make a few remarks. When used as models for galaxies, general relativistic dust systems are not intended to represent entire galaxies, but rather specific regions where the dust approximation and symmetry assumptions hold, typically areas near the galactic plane and sufficiently far from the rotation axis. The notion that a single metric could describe a whole galaxy is, in fact, somewhat naive. Moreover, the symmetries invoked have inherent limitations. As emphasized by us \cite{Astesiano:2022ghr} and \citet{Galoppo:2024mfw}, the assumption of stationarity restricts the validity of the solution to timescales short relative to those of galaxy formation and evolution. Similarly, the imposition of axial symmetry limits its applicability to timescales longer than those associated with the effective dust-fluid approximation of the time-averaged oscillatory motion of stars and gas in the galactic plane. Consequently, the model is valid over a well-defined temporal range, which can be estimated as approximately  $10^{7}  \mathrm{yr}  \lesssim t  \lesssim 10^{9} \mathrm{yr}$ \cite{Galoppo:2024mfw}. 
Another relevant issue is that of the stability of these systems on the relevant time scales: in this case, we know that pressure terms are related to the velocity dispersion in disc galaxies, they support the thickness of the disc, and are relevant for its stability. Neglecting pressure, thus, perhaps oversimplifies the model (see \citet{Galoppo:2024ttc} and references therein). Nonetheless, despite these limitations, the model under consideration exhibits distinctive non-Newtonian features that may have significant implications on large scales, such as those of galaxies \cite{cacciatori}. Probably, these features are related to the problem of the impact of the large-scale physics and geometry on the local physical systems, which is very well known in cosmology \cite{ellis}.
The main strength of this model lies in its capacity to investigate non-Newtonian effects that emerge in the low-energy limit. Once these features are clarified, more complex and realistic models can be developed which, presumably, will retain its distinctive characteristics. }

Here, we focused on gravitational lensing observables arising from rotation effects, which are a distinctive feature of these systems. We showed that the impact of the dragging terms is threefold. First, the presence of $\psi$ modifies the relationship between the sources of the gravitational field and the Newtonian potential $\Phi$, as demonstrated by the modified Poisson equation (\ref{eq:poisson}). Second, the dragging terms appear not only in the off-diagonal components of the metric (\ref{eq:metricsol}) but also in the diagonal ones. As discussed in \citet{Astesiano:2024zzz}, a scalar-vector-tensor decomposition of the metric reveals that the scalar degree of freedom includes both $\Phi$ and $\psi$, unlike in the solution to Einstein's equations for isolated sources in the weak-field limit, where scalar and vector perturbations remain distinct. Eventually, the dragging terms have an impact on the gravitomagnetic contribution to the deflection angle (\ref{eq:defalphaGM1}). 

{The first two of the effects listed are quadratic in $\psi$ and  give a contribution which is  $O(c^{-2})$ which can be interpreted as an additional effective mass. Our analysis show that purely geometrical effects results in additional effective mass, and we think that their presence should be taken into account to properly model the presence of dark matter. If we suppose that on these scales $\displaystyle \psi \simeq V_{\psi} L_{\psi}$, where $V_{\psi}, L_{\psi}$ denote reference velocity and length of the rotating dust model,  the extra mass contribution $M_{\psi}$ can be estimated as $\displaystyle GM_{\psi} \simeq \frac{ V^{2}_{\psi} L^{2}_{\psi}}{R}$, where $R$ is the length scale of lensing.} 

The gravitomagnetic contribution to the deflection angle (\ref{eq:defalphaGM1}) exhibits additional distinctive features that reflect the symmetries inherent to its vectorial nature. Gravitomagnetic corrections to gravitational lensing (as for lensing in spherically symmetric spacetimes, see e.g. \citet{kv1,kv2} and references therein) have been thoroughly investigated in previous studies, as evidenced by numerous works \cite{sereno02,sereno03MNRAS,serenoPRD2003,2005MNRAS.357.1205S,rotation1,rotation2,rotation3,rotation4,rotation5bozza,rotation6}, which considered both compact sources, such as stars, and extended ones, such as galaxies and focused on also other observables, such as the time delay and the gravitational Faraday rotation\cite{faraday_ruggiero}. However, all of these studies were carried out within the weak-field approximation of Einstein's equations, where the gravitomagnetic terms are of order $O(c^{-3})$ and thus significantly smaller than the leading contributions from the Newtonian potential. In stark contrast, the self-gravitating rotating systems examined in this work feature dragging effects that produce gravitomagnetic contributions of order $O(c^{-1})$, representing a fundamental departure from the standard weak-field scenario: in fact, we referred to them as \textit{strong gravitomagnetims} \cite{Astesiano:2024zzz}. 
In terms of the extra mass contribution $M_{\psi}$ and the lensing length scale $R$ discussed above, the leading gravitomagnetic contribution can be written as $ \displaystyle \alpha_{GM} \simeq \sqrt{\frac{GM_{\psi}}{c^{2}R}}$. Thus, a relation between the observed asymmetry and the matter content of geometric origin can be established. {In addition, a detection of this contribution can constrain the quasilocal angular momentum of the system $\psi$; from dimensional analysis, we get in fact $\displaystyle \psi \simeq 4.5 \times 10^{19 }\left(\frac{\alpha_{GM}}{\mathrm{1\, mas}}\right)\left(\frac{R}{1\, \mathrm{kpc}}\right)$, where a resolution of milliarcsecond was assumed for the lensing observables.}

To compute the effects explicitly, we selected a specific solution of the Grad-Shafranov equation (\ref{eq:GS1}), because the spacetime metric is not uniquely fixed by the density and velocity profiles of the sources. {As previously discussed, the expression for $\psi$ in Eq. (\ref{eq:psidef1}), which is subsequently employed in the illustrative examples, corresponds to the dipole contribution arising in the weak-field expansion of the Kerr solution. We note that, although making quantitative predictions about the magnitude of this kind of effects requires selecting a specific solution of the Grad–Shafranov equation, the asymmetry observed in image formation is independent of the particular form of $\psi$, as it stems from the inherently vectorial nature of the frame-dragging effects. On the other hand, as discussed above, the presence of these effects can be inferred indirectly as an effective additional matter, requiring a shift in perspective from a particle-based interpretation to one framed in terms of rotational energy.  } One might ask where the solutions of the Grad-Shafranov equation originate. The idea is that they are tied to the system’s symmetry and its temporal evolution. Although we are considering a stationary system by definition, this stationarity is only approximate, justified because the observation time is short compared to the system’s age. Therefore, it is plausible that these solutions reflect the system's history. As a simple illustration, consider the line element $\displaystyle ds^{2}=-\left(1-{\Omega^{2}r^{2}}{} \right)dt^{2}+2{\Omega r^{2}}{}dtd\varphi +dr^{2}+r^{2}d\varphi^{2}+dz^{2} $, which describes a frame rotating with angular velocity $\Omega$ in Minkowski spacetime \cite{Rizzi:2002sk}. If, however, we imagine starting from a rest frame and accelerating it into rotation, then $\Omega=\alpha \tau$ would depend on the constant angular acceleration $\alpha$ and the duration $\tau$ over which it is applied. In this sense, the function $\psi$ can be thought of as a remnant of the system’s dynamical past.

In conclusion, the effects discussed here are not entirely new; rather, the novelty lies in the idea that, if such relativistic rotating dust systems exist, their presence could be revealed through these effects, which are not negligible. In other words, we propose that observed asymmetries in gravitational lensing phenomena may originate from the existence of these systems. Conversely, if the presence of such dust configurations is confirmed, their influence on dark matter tracers, such as rotation curves and gravitational lensing, should be properly accounted for.

\section*{Acknowledgements}

The authors thank Marco Galoppo and Kumar Shwetketu Virbhadra for their suggestions and valuable insights. M.L.R. gives thanks for the support of the Gruppo Nazionale per la Fisica Matematica (GNFM).

\bibliography{refs_lensing}

\end{document}